\documentclass[twocolumn]{autart}   
\usepackage{graphicx}         
\usepackage{xcolor}
\usepackage{comment}
\usepackage{float}
\usepackage{lipsum}

\usepackage{enumitem}
\setlist[enumerate]{leftmargin=*}
\newtheorem{assumption}{Assumption}

\newtheorem{lemma}{Lemma}
\newtheorem{proposition}{Proposition}

\newtheorem{example}{Example}

\newtheorem{remark}{Remark}
\usepackage{comment}
\usepackage{algorithmicx}
\usepackage{amsmath}
\usepackage{amssymb}
\usepackage{mathtools}

\DeclareMathOperator{\diag}{diag}

\DeclareMathOperator{\Id}{Id}
\newcommand{\at}{\tilde{A}}
\newcommand{\bt}{\tilde{B}}

\newcommand{\OO}{\mathcal{O}}
\newcommand{\RR}{\mathbb{R}}
\newcommand{\NN}{\mathbb{N}}

\usepackage{bm}
\usepackage{algorithm}
\usepackage{algpseudocode}

\begin{document}

\begin{frontmatter}
\title{Systematic interval observer design for linear systems}
\thanks{This research was supported in part by a fellowship by the Hanse-Wissenschaftskolleg---Institute for Advanced Study.}
\thanks[footnoteinfo]{The authors contributed equally to this work.}
%Their names are listed in alphabetical order. Corresponding author T.~N.~Dinh. Tel. +33 1 40 27 25 90.}

\author[Paestum]{Thach Ngoc Dinh\thanksref{footnoteinfo}}\ead{ngoc-thach.dinh@lecnam.net},   
\author[Rome]{Gia Quoc Bao Tran\thanksref{footnoteinfo}}\ead{gia-quoc-bao.tran@minesparis.psl.eu}        

\address[Paestum]{Conservatoire National des Arts et M\'etiers (CNAM), Cedric-Lab, 292 rue St-Martin, 75141 Paris Cedex 03, France}       
\address[Rome]{Centre Automatique et Syst\`emes (CAS), Mines Paris, Universit\'e PSL, Paris, France}
          

\begin{abstract}
We first develop systematic and comprehensive interval observer designs for linear time-invariant (LTI) systems, under standard assumptions of observability and interval bounds on the initial condition and uncertainties. Traditionally, such designs rely on specific transformations into Metzler (in continuous time) or non-negative (in discrete time) forms, which may impose limitations. We demonstrate that these can be effectively replaced by an LTI transformation that is straightforward to compute offline. Subsequently, we extend the framework to time-varying systems, overcoming the limitations of conventional approaches that offer no guarantees. Our method utilizes dynamic transformations into higher-dimensional target systems, for which interval observers can always be constructed. These transformations become left-invertible after a finite time, provided the system is observable and the target dynamics are sufficiently high-dimensional and fast, thereby enabling the finite-time recovery of interval bounds in the original coordinates. Academic examples are provided to illustrate the proposed methodology.
\end{abstract}

\end{frontmatter}

\section{Introduction}
To begin this paper, we present the continuous-time (CT) and discrete-time (DT) cases in a unified manner, as their structures are similar. Let $x_t$ denote the current state, encompassing the conventional $x(t)$ in CT and $x_k$ in DT. Likewise, $x_t^+$ represents the time derivative or the forward shift of the state, i.e., $\dot{x}(t)$ in CT and $x_{k+1}$ in DT. Consider the linear time-invariant (LTI) system:
\begin{equation}\label{eq:sysxlin}
    x_t^+ = F x_t + u_t + Dd_t, \quad
    y_t = H x_t + Ww_t,
\end{equation}
where $x_t \in \mathbb{R}^{n_x}$ is the state, $u_t \in \mathbb{R}^{n_x}$ is the known control input, and $y_t \in \mathbb{R}^{n_y}$ is the measured output. The vectors $(d_t,w_t) \in \mathbb{R}^{n_d} \times \mathbb{R}^{n_w}$ represent the unknown input disturbance and measurement noise. The matrices $(F,H,D,W)$ are known and constant. To design for system~\eqref{eq:sysxlin} an \emph{interval observer} (see~\cite[Definition 1]{Mazenccontinu} for CT and~\cite[Definition 1]{mazenc2014interval} for DT), existing methods require that $F$ be Metzler$\slash$cooperative (in CT) or non-negative (in DT). But this may not inherently be the case, so we typically employ a transformation into the desired form. The first Jordan form-based transformation we may use is~\cite{Mazenccontinu} for CT and~\cite{mazenc2014interval} for DT, for which we need to deploy a \emph{time-varying} $n_x \times n_x$ transformation, leading to an undesired time-varying observer for a time-invariant system. While this transformation exists for all constant real matrices, its closed form is cumbersome to compute in higher dimensions; moreover, at each time step, we need to update $2n_x^2$ values (twice for the forward and inverse computations), potentially causing a significant computational burden. Furthermore, there is no freedom in the choice of the target form. Indeed, given $F$, we rely on~\cite{Mazenccontinu,mazenc2014interval} to compute the transformation, then the target dynamics matrix follows this without any freedom. Last, the difference between CT~\cite{Mazenccontinu} and DT~\cite{mazenc2014interval} results is significant, rendering this concept unsystematic for LTI systems. We may consider an alternative change of coordinates as proposed by~\cite{Raissietal,Efimov-12}, which has been crafted for both CT and DT systems. However, this approach necessitates, without giving constructive guarantees, the existence of \emph{two additional vectors} that, in conjunction with the stable matrices in both the original and target coordinates, form two observable pairs~\cite[Lemma 1]{Raissietal}. Consequently, the flexibility of the target form, though more realistic than~\cite{Mazenccontinu,mazenc2014interval}, is still somewhat constrained.

In the reviewed methods, the interval observer involves two key design parameters: the coordinate transformation $P$ and the gain $L$. Their interdependence obstructs a unified transformation scheme for linear systems, motivating this paper where we propose a systematic interval observer design for system~\eqref{eq:sysxlin} with a simple built-in transformation based on~\cite{luenberger}. This allows decoupling the transformation from the observer design, thus breaking the mentioned interdependence. Our design offers significant advantages compared to the mentioned methods. First, both the transformation and observer design follow a systematic process. Unlike methods that rely on interdependent matrices $(P,L)$, we require only a single \emph{constant} transformation matrix $T$, obtained offline by solving a Sylvester equation with guaranteed solution existence. This results in a \emph{time-invariant} observer and offers significant computational efficiency compared to~\cite{Mazenccontinu,mazenc2014interval,Raissietal,Efimov-12}. Second, it lets us almost freely specify the target form, which is (almost arbitrarily) chosen \emph{before} the transformation is fixed, unlike in~\cite{Mazenccontinu,mazenc2014interval}. Importantly, our method, using the Kravaris-Kazantzis$\slash$Luenberger (KKL) framework, extends to linear time-varying (LTV) systems:
\begin{equation}\label{eq:sysxltvintro}
        x_t^+ = F_t x_t + u_t+ D_td_t, \quad
        y_t = H_tx_t + W_tw_t.
\end{equation}
This framework involves transforming into higher-dimensional stable dynamics for observer design and inverting the transformation---see~\cite{bernardNonautonomousKKL,baokkltac}.
To the best of our knowledge, there have not yet been any such systematic designs for LTV systems. 
Leveraging the inherent universality of the KKL paradigm, extending our methodology as in Section~\ref{sec_ltv} requires minimal effort, unlike extending~\cite{Mazenccontinu,mazenc2014interval} to LTV systems by re-computing online Jordan-based transformations. Note that~\cite{Raissietal,gu} also employ Sylvester equations for interval observer design in LTI systems without extending to LTV systems, while~\cite{Efimov-12} made similar attempts with less straightforward assumptions, as discussed in detail in Section~\ref{sec_ltv}.

\emph{Notations:}
Denote $\NN = \{0,1,\ldots\}$ and $\NN_{\geq m} = \{m,m+1,\ldots\}$ for $m \in \NN$. Inequalities like $a \leq b$ for vectors $a$, $b$ or $A \leq B$ for matrices $A$, $B$ are component-wise. For a matrix $M \in \RR^{m \times n}$ with entries $m_{i,j}$, let $M^\oplus$ be the matrix in $\RR^{m \times n}$ with entries $\max\left\{0, m_{i, j}\right\}$ and $M^\ominus = M^\oplus - M$. Let $M^\dagger$ be the Moore-Penrose inverse of matrix $M$ and $M^{-1}$ be the inverse of invertible square matrix $M$. Let $\diag(\lambda_1,\lambda_2,\ldots,\lambda_n)$ and $\Id$ be the diagonal matrix and the identity matrix. Let $\|\cdot\|$ be a matrix norm.

\begin{lemma}\cite[Section II.A]{Efimov-12}
\label{lem_pm}
Consider vectors $a$, $\overline a$, $\underline a$ in $\mathbb{R}^{n}$ such that $\underline a \leq a \leq \overline a$. For any $A \in \RR^{m\times n}$, we have $A^{\oplus}\underline a -A^{\ominus}\overline a \leq Aa \leq A^{\oplus}\overline a -A^{\ominus}\underline a$.
\end{lemma}

\section{LTI interval observers for LTI systems}\label{sec_lin}
Consider system~\eqref{eq:sysxlin} with these standard assumptions.
\begin{assumption}\label{ass_syslin}
    For system~\eqref{eq:sysxlin}, we assume that:
\begin{enumerate}[label=(A1.\arabic*),leftmargin=*,nosep]
        \item \label{ass_obslin} The pair $(F,H)$ is observable;
        \item \label{ass_x0lin} There exist $\overline x_0$ and $\underline x_0$ such that $\underline x_0 \leq x_0 \leq \overline x_0$;
        \item \label{ass_dwlin} There exist known vectors $(\overline d_t,\underline d_t,\overline w_t,\underline w_t)$ such that $\underline d_t \leq d_t \leq \overline d_t$ and $\underline w_t \leq w_t \leq \overline w_t$ for all $t \geq 0$.
    \end{enumerate}
\end{assumption}

Under Assumption~\ref{ass_syslin}, inspired by~\cite[Section 3.1]{franklin}, we propose for system~\eqref{eq:sysxlin} the interval observer
\begin{subequations}\label{eq:obszlin}
\begin{equation}
\left\{ \begin{array}{@{}r@{\;}c@{\;}l@{}}
        \overline z_t^+ &=& A \overline z_t + B y_t + Tu_t + (TD)^\oplus \overline d_t -(TD)^\ominus \underline d_t\\&&{}   + (BW)^\ominus \overline w_t - (BW)^\oplus \underline w_t\\
        \underline z_t^+&=& A \underline z_t + B y_t + Tu_t + (TD)^\oplus \underline d_t -(TD)^\ominus \overline d_t\\&&{} + (BW)^\ominus \underline w_t - (BW)^\oplus \overline w_t,
    \end{array}\right.
\end{equation}
with states $\overline{z}_t$, $\underline{z}_t$ in $\RR^{n_x}$, with the initial conditions
\begin{equation}\label{eq:initlin}
\overline z_0= T^\oplus \overline x_0 - T^\ominus \underline x_0,\quad
        \underline z_0 =  T^\oplus \underline x_0 - T^\ominus \overline x_0,
\end{equation}
and the bounds of $x_t$ at all times after time $0$ given by
\begin{align}\label{eq:backp}
    \overline x_t &= (T^{-1})^\oplus \overline z_t - (T^{-1})^\ominus \underline z_t, \\\label{eq:backm}
    \underline x_t &= (T^{-1})^\oplus \underline z_t-(T^{-1})^\ominus \overline z_t,
\end{align}
\end{subequations}
where $T \in \RR^{n_x \times n_x}$ is solution to the Sylvester equation
\begin{equation}\label{eq:sylvesterlin}
    TF = AT + BH,
\end{equation}
with $(A,B)$ to be defined. While the survey~\cite[Section 3.1]{franklin} focuses exclusively on CT LTI systems based on~\cite{luenberger}, the analogous result applies to DT systems and is thus presented herein. The following proposition, formulated based on~\cite{franklin} for both CT and DT contexts without the necessity of introducing the gain matrix $L$, is stated.
\begin{proposition}\label{lem_linearcase}
    Suppose Assumption~\ref{ass_syslin} holds. For any $A \in \RR^{n_x \times n_x}$ either Hurwitz and Metzler in CT, or Schur and non-negative in DT, with eigenvalues different from those of $F$, and for any $B \in \RR^{n_x \times n_y}$ such that $(A,B)$ is controllable, if $T$ solution to~\eqref{eq:sylvesterlin} is invertible, then
    observer~\eqref{eq:obszlin} is an interval observer for system~\eqref{eq:sysxlin}.
\end{proposition}

\textbf{Proof.} Pick $(A,B)$ as in Proposition~\ref{lem_linearcase}. From the conditions stated in Proposition~\ref{lem_linearcase} and from Remark~\ref{rem_inv}, there exists a unique invertible solution $T$ to~\eqref{eq:sylvesterlin}. Now, we show that observer~\eqref{eq:obszlin} is an interval observer for system~\eqref{eq:sysxlin}. Since $T$ satisfies~\eqref{eq:sylvesterlin}, the variable $z_t := T x_t$ is solution to the dynamics $z_t^+ = A z_t + B y_t + Tu_t + TDd_t - BWw_t$.
From Item~\ref{ass_x0lin} of Assumption~\ref{ass_syslin} and $z_0=Tx_0$, we deduce that $\underline z_0\leq z_0 \leq \overline z_0$, where $\underline z_0$ and $\overline z_0$ are given in~\eqref{eq:initlin}. Consider the solutions $(z_t,\overline z_t,\underline z_t)_{t \geq 0}$ to
\begin{equation}\label{eq:extensions}
\left\{ \begin{array}{@{}r@{\;}c@{\;}l@{}}
        z_t^+ &=& A z_t + B y_t + Tu_t + TDd_t - BWw_t\\
        \overline z_t^+ &=& A\overline z_t + B y_t + Tu_t  + (TD)^\oplus \overline d_t -(TD)^\ominus \underline d_t \\&&{}+ (BW)^\ominus \overline w_t - (BW)^\oplus \underline w_t\\
        \underline z_t^+&=& A\underline z_t + B y_t + Tu_t +(TD)^\oplus \underline d_t -(TD)^\ominus \overline d_t\\&&{} + (BW)^\ominus \underline w_t - (BW)^\oplus \overline w_t.
    \end{array}\right.
\end{equation}
 We get
 \begin{align*}
 \overline z_t^+-z_t^+ &= A(\overline z_t-z_t) 
       +(TD)^\oplus \overline d_t -(TD)^\ominus \underline d_t \\&\qquad{}-TDd_t +(BW)^\ominus \overline w_t - (BW)^\oplus \underline w_t+Bw_t,\\
       z_t^+-\underline z_t^+&= A(z_t-\underline z_t) +TDd_t -((TD)^\oplus \underline d_t -(TD)^\ominus \overline d_t ) \\&\qquad{} - (BW)^\ominus \underline w_t + (BW)^\oplus \overline w_t-Bw_t.
 \end{align*}
    From Lemma~\ref{lem_pm}, we get for all $t \geq 0$,
    \begin{align*}
    (TD)^\oplus \underline d_t - (TD)^\ominus \overline d_t &\leq TDd_t 
    \\&\leq (TD)^\oplus \overline d_t - (TD)^\ominus \underline d_t,\\
    (BW)^\ominus \overline w_t -(BW)^\oplus \underline w_t &\geq -BWw_t 
    \\&\geq (BW)^\ominus \underline w_t-(BW)^\oplus \overline w_t.
    \end{align*}
    Hence, we get 
    \begin{align*}
    p_t&=(TD)^\oplus \overline d_t -(TD)^\ominus \underline d_t -TDd_t +(BW)^\ominus \overline w_t \\&\qquad{}- (BW)^\oplus \underline w_t+BWw_t \geq 0,\\
    q_t&=TDd_t -((TD)^\oplus \underline d_t -(TD)^\ominus \overline d_t ) - (BW)^\ominus \underline w_t \\&\qquad{}+ (BW)^\oplus \overline w_t-BWw_t\geq 0.
    \end{align*}
    Because the matrix $A$ is Metzler in CT or non-negative in DT, $p_t \geq 0$ and $q_t \geq 0$ for all $t \geq 0$, and $0 \leq \overline z_0-z_0$ and $0 \leq z_0 -\underline z_0$, we can deduce that $\underline z_t \leq z_t:= Tx_t \leq \overline z_t$ for all $t \geq 0$. Thus, from~\eqref{eq:backp},~\eqref{eq:backm}, and Lemma~\ref{lem_pm}, we conclude that $\underline x_t \leq x_t \leq \overline x_t$ for all $t \geq 0$. Finally, we deduce from system~\eqref{eq:extensions} that, in the absence of $d_t$ and $w_t$, $\overline z_t^+-\underline z_t^+=A(\overline z_t-\underline z_t)$.
Since $A$ is also Hurwitz in CT or Schur in DT, we have $\displaystyle\lim_{t\to+\infty}(\overline z_t-\underline z_t) = 0$. Thus, from~\eqref{eq:backp} and~\eqref{eq:backm}, we get $\displaystyle \lim_{t\to+\infty}(\overline x_t-\underline x_t)=\displaystyle \lim_{t\to+\infty}(\overline z_t-\underline z_t)=0$. \hfill $\blacksquare$

\begin{remark}\label{rem_inv}
If $n_y = 1$,~\cite{luenberger} shows that $T$ solution to~\eqref{eq:sylvesterlin} is invertible under the conditions stated in Proposition~\ref{lem_linearcase}. Thus, the invertibility condition in Proposition~\ref{lem_linearcase} is inherently satisfied under observability and when $(A,B)$ is controllable. When $n_y \in \NN_{\geq 2}$, we can show that this invertibility holds under observability of $(F,H)$ for \emph{almost any} pair $(A,B)$ (note that almost any pair $(A,B)$ is controllable), i.e., the set of $(A,B)$ making $T$ non-invertible has zero Lebesgue measure in the space of real matrices. Indeed, from~\cite{hucheng}, $T$ is rational in the entries of $A$ and $B$.\footnote{Note: $\gamma_0$ in~\cite[Eq.~19]{hucheng} is implicitly a polynomial in $(A,B)$.} It follows that the determinant of $T$ is also rational in $(A,B)$ and that $T$ being invertible is equivalent to a non-zero polynomial of the entries of $(A,B)$. Moreover, due to the observability of $(F,H)$, there exists $(A,B)$ making this polynomial non-zero. Combining this with~\cite{caronrichard}, we get that $T$ is invertible for almost any $(A,B)$. Therefore, Proposition~\ref{lem_linearcase} is generic, and $T$ is almost always invertible under Assumption~\ref{ass_syslin} for $n_y \in \NN_{\geq 2}$.
\end{remark}

Exploiting system~\eqref{eq:sysxlin} being LTI, we can also write and implement directly the observer in the $x$-coordinates as in~\cite{thachecc}, obtaining the following form
\begin{subequations}\label{eq:obsxlin}
\begin{equation}\label{eq:exten}
 \left\{ \begin{array}{@{}r@{\;}c@{\;}l@{}}
        \overline{\hat{x}}^+_t &=& F \overline{\hat{x}}_t + u_t + T^{-1}B (y_t - H \overline{\hat{x}}_t)  \\&&{}+ T^{-1} ((TD)^\oplus \overline d_t - (TD)^\ominus \underline d_t) \\&&{} + T^{-1} ((BW)^\ominus \overline w_t - (BW)^\oplus \underline w_t)\\
       \underline{\hat{x}}^+_t&=& F \underline{\hat{x}}_t + u_t + T^{-1}B (y_t - H \underline{\hat{x}}_t) \\&&{}+ T^{-1}((TD)^\oplus \underline d_t - (TD)^\ominus \overline d_t) \\&&{} + T^{-1}((BW)^\ominus \underline w_t - (BW)^\oplus \overline w_t),
    \end{array}\right.
\end{equation}
associated with the initial conditions
\begin{align}\label{eq:initxlinp}
        \overline{\hat{x}}_0 &= T^{-1}(T^\oplus \overline x_0-T^\ominus \underline x_0),\\\label{eq:initxlinm}
        \underline{\hat{x}}_0 &= T^{-1}(T^\oplus \underline x_0-T^\ominus \overline x_0),
\end{align}
and the bounds after time $0$
\begin{align}\label{eq:initbxlinp} 
      \overline x_t&= (T^{-1})^\oplus T\overline{\hat{x}}_t- (T^{-1})^\ominus T\underline{\hat{x}}_t, \\\label{eq:initbxlinm}
        \underline x_t&= (T^{-1})^\oplus T\underline{\hat{x}}_t - (T^{-1})^\ominus T\overline{\hat{x}}_t.
\end{align}
\end{subequations}
The dependence of observer~\eqref{eq:obsxlin} on the chosen $A$ lies implicitly in $T$, which is solution to~\eqref{eq:sylvesterlin}. We state the next proposition, whose proof resembles that of Proposition~\ref{lem_linearcase}.
\begin{proposition}\label{lem_linearcase2}
    Suppose Assumption~\ref{ass_syslin} holds. 
    With the choice of $(A,B)$ stated in Proposition~\ref{lem_linearcase}, observer~\eqref{eq:obsxlin} with $T$ solution to~\eqref{eq:sylvesterlin}, if $T$ is invertible (see Remark~\ref{rem_inv}), is an interval observer for system~\eqref{eq:sysxlin}.
\end{proposition}

Note that observers~\eqref{eq:obszlin} and~\eqref{eq:obsxlin} are time-invariant, which differs from the unnecessarily time-varying approach for LTI systems in~\cite{Mazenccontinu,mazenc2014interval}. We can now state a novel, unified framework for interval observer design in LTI systems.
\begin{algorithm}
\caption{Interval observer design for system~\eqref{eq:sysxlin}}\label{alg:cap}
\begin{algorithmic}
\Require System~\eqref{eq:sysxlin} under Assumption~\ref{ass_syslin}
\State \textbf{Step 1:} Pick $(A,B)$ satisfying Proposition~\ref{lem_linearcase} or~\ref{lem_linearcase2}
\State \textbf{Step 2:} Solve~\eqref{eq:sylvesterlin} one time for $T$  \Comment{In Matlab, simply use {\fontfamily{qcr}\selectfont T = sylvester(A,-F,-B*H)}}
\State \textbf{Step 3:} Implement observer~\eqref{eq:obszlin} or~\eqref{eq:obsxlin}
\end{algorithmic}
\end{algorithm}

The following academic example illustrates our method.
\begin{example}
    To illustrate Algorithm~\ref{alg:cap} in handling high-dimensional systems compared to existing methods, we randomly choose an observable eighth-dimensional LTI CT system with $F\in\RR^{8\times 8}$, $D\in\RR^{8\times 2}$, $H\in\RR^{6\times 8}$, and $W\in\RR^{6\times 1}$ utilizing the {\fontfamily{qcr}\selectfont randi} command in Matlab. While conventional transformations in~\cite{Mazenccontinu,mazenc2014interval,Raissietal,Efimov-12} may result in less flexibility in selecting the target form or impose significant computational burdens for large-dimensional systems, our approach in Algorithm~\ref{alg:cap} proves to be very straightforward in this context. Pick $(A,B)$ as in Proposition~\ref{lem_linearcase} or~\ref{lem_linearcase2} and compute $T$. Consider known $u_t=(1,1,\ldots,1)\in\RR^8$, unknown $d_t = 0.5(\sin(2t),\cos(t))$ and $w_t = 0.3\sin(t)$ with some known bounds, and simulate observer~\eqref{eq:obsxlin}. Estimation results are shown in Figure~\ref{fig1}. 
    \begin{figure}%[H]
        \centering
\includegraphics[width=0.5\textwidth,height=0.234\textwidth]{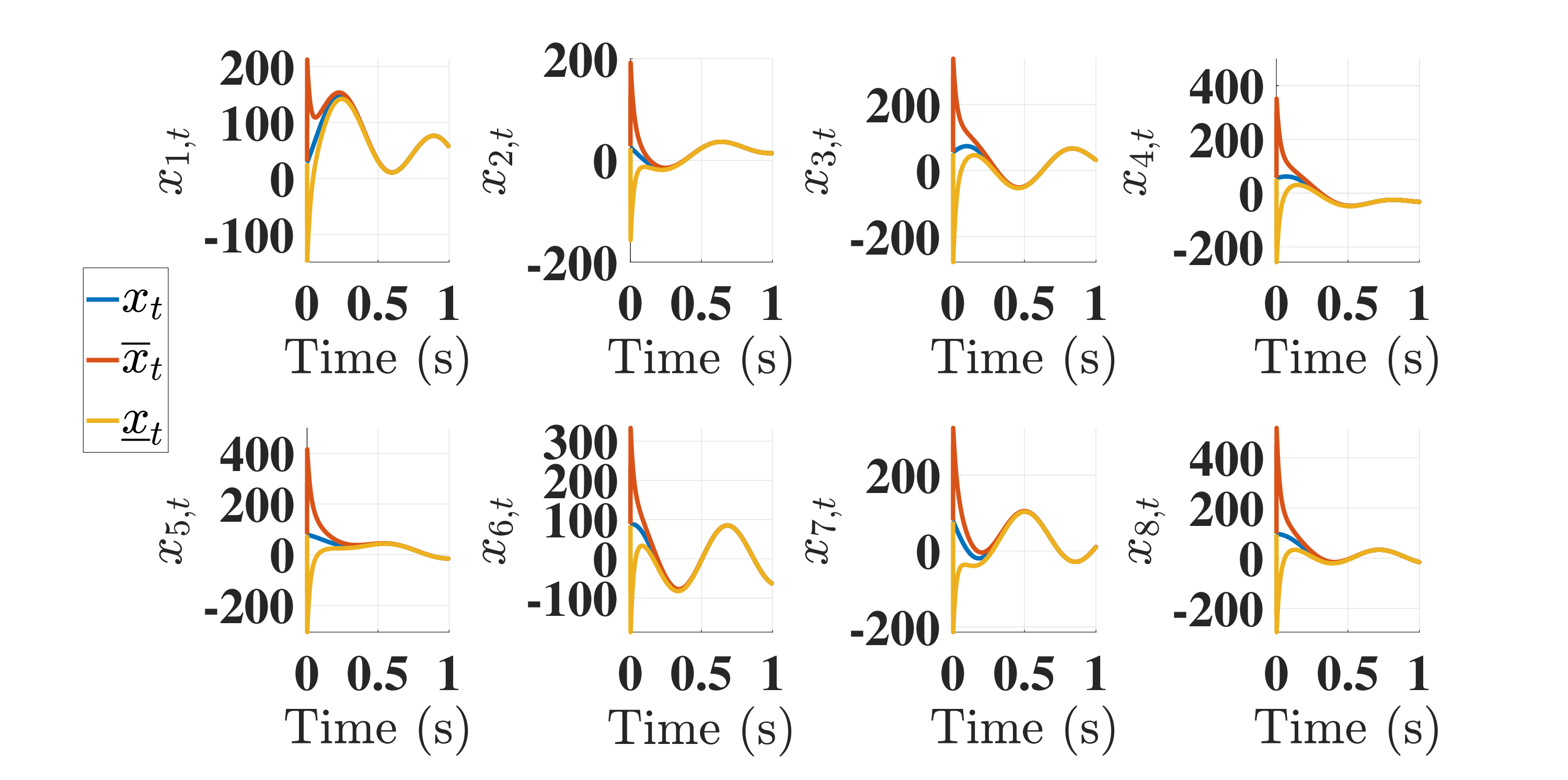}
\includegraphics[width=0.5\textwidth,height=0.234\textwidth]{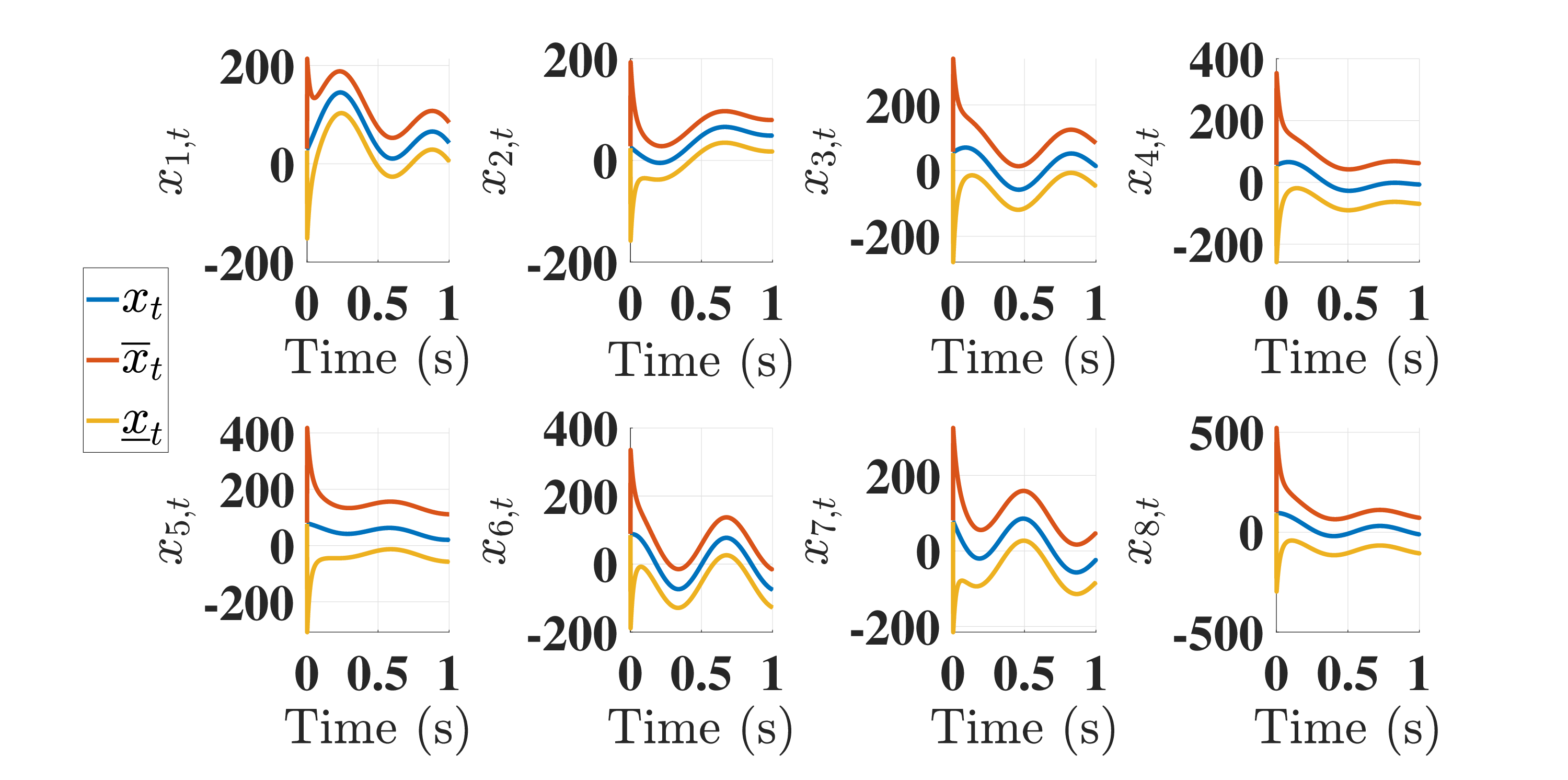}
\caption{Estimation of $x_{t} \in \RR^8$ in the absence and presence of $(d_t,w_t)$, above and below respectively, using observer~\eqref{eq:obsxlin}.}       \label{fig1}
    \end{figure}
\end{example}

\section{LTV interval observers for LTV systems}\label{sec_ltv}
In this section, our focus is on LTV systems. First, it is important to recognize that finding a Jordan form for a time-varying matrix at each instant based on~\cite{Mazenccontinu,mazenc2014interval} 
is not realistic because the transformation's closed form must be computed again at each time, thus the effectiveness of establishing an LTV change of coordinates $P(t)$ (resp., $(R_k)_{k \in \NN}$) through the Jordan form as in~\cite{Mazenccontinu} (resp.,~\cite{mazenc2014interval}) is very limited, especially if the matrix changes rapidly over time. However, our approach can easily deal with this matter through systematic KKL-based LTV transformations. Second, compared to~\cite{Efimov-12}, while our change of coordinates is time-varying, the target system~\eqref{eq:syszltv_d} is time-invariant in the dynamics part. This enables us to circumvent the necessity of requiring a common Lyapunov function for all time steps $k$ as in~\cite[Assumption 4]{Efimov-12}, and eliminates the need for the restrictive assumption regarding the existence of a time-invariant change of coordinates for time-varying systems as in~\cite[Assumption 5]{Efimov-12}. Since KKL results differ between CT and DT in the LTV setting, let us consider each case separately.

\subsection{Case of CT LTV systems}
In this section, consider a CT LTV system
\begin{subequations}\label{eq:sysxltv_c}
\begin{align}
        \dot{x}(t) &= F(t) x(t) + u(t) + D(t)d(t), \\
        y(t) &= H(t) x(t) + W(t)w(t),
\end{align}
\end{subequations}
with $(x(t),u(t),y(t),d(t),w(t))$ defined in system~\eqref{eq:sysxltvintro} and $(F(t),H(t),D(t),W(t))$ known matrices.
\begin{assumption}\label{ass_sysltv_c}
    For system~\eqref{eq:sysxltv_c}, we assume that:
\begin{enumerate}[label=(A2.\arabic*),leftmargin=*,nosep]
        \item \label{ass_fhltv_c} There exist $c_f > 0$ and $c_h > 0$ such that for all $t \geq 0$, we have $\|F(t)\| \leq c_f$ and $\|H(t)\| \leq c_h$;
        \item \label{ass_obsltv_c} The pair $(F(t),H(t))$ is instantaneously observable,\footnote{This is also known as \emph{differential observability}, namely the output and a finite number of its derivatives determine uniquely the solution. See~\cite[Section II.B]{BriPetPra} for more details.} i.e., there exist $c_o > 0$ and for each $i \in \{1,2,\ldots,n_y\}$, $m_i \in \NN$, such that for all $t \geq 0$, we have
        $
\sum_{i = 1}^{n_y} (\OO_i(t))^\top \OO_i(t) \geq c_o \Id
$,
where $\OO_i(t) = (\OO_{i,1}(t),\OO_{i,2}(t),\ldots,\OO_{i,m_i}(t))$ where $\OO_{i,1}(t) = H_i(t)$ and $\OO_{i,j+1}(t) = \dot{\OO}_{i,j}(t) + \OO_{i,j}(t)F(t)$, with $H_i(t)$ the $i^{\rm th}$ row of $H(t)$;
        \item \label{ass_x0ltv_c} There exist $\underline x_0$ and $\overline x_0$ such that $\underline x_0 \leq x(0) \leq \overline x_0$;
        \item \label{ass_dwltv_c} There exist known vectors $(\overline d(t),\underline d(t),\overline w(t),\underline w(t))$ such that $\underline d(t) \leq d(t) \leq \overline d(t)$ and $\underline w(t) \leq w(t) \leq \overline w(t)$ for all $t \geq 0$.
    \end{enumerate}
\end{assumption}
Following the KKL paradigm~\cite{bernardNonautonomousKKL}, we strive for an LTV transformation $x(t) \mapsto z(t) := T(t) x(t) \in \RR^{n_z}$, where $t \mapsto T(t) \in \RR^{n_z \times n_x}$, for some $n_z \geq n_x$ defined later, is a time-varying matrix satisfying
\begin{equation}\label{eq:sylvesterltv_c}
    \dot{T}(t) + T(t)F(t) = AT(t) + BH(t), \quad \forall t \geq 0,
\end{equation}
where $A$ is Hurwitz and Metzler, and $(A,B)$ is controllable. As a result, system~\eqref{eq:sysxltv_c} is put into
\begin{multline}\label{eq:syszltv_c}
        \dot{z}(t) = A z(t) + B y(t) + T(t) u(t) \\+ T(t) D(t) d(t) - BW(t)w(t).
\end{multline}
We propose the interval observer for system~\eqref{eq:syszltv_c}:
\begin{subequations}\label{eq:obszltv_c}
\begin{equation}
\left\{\begin{array}{@{}r@{\;}c@{\;}l@{}}\label{eq:intobzltv_c}
        \dot{\overline{z}}(t) &=&A \overline{z}(t) +B y(t) + T(t) u(t) \\&&{} + (T(t)D(t))^\oplus \overline d(t) -(T(t)D(t))^\ominus \underline d(t) \\&&{}+ (BW(t))^\ominus \overline w(t) - (BW(t))^\oplus \underline w(t)\\
        \dot{\underline{z}}(t) &=& A \underline{z}(t) + B y(t) + T(t) u(t) \\&&{}+(T(t)D(t))^\oplus \underline d(t) -(T(t)D(t))^\ominus \overline d(t) \\&&{}+ (BW(t))^\ominus \underline w(t) - (BW(t))^\oplus \overline w(t),
    \end{array}\right.
\end{equation}
where the dynamics of $T$ can be updated as solution to~\eqref{eq:sylvesterltv_c} as
\begin{equation}\label{eq:sylltv_c}
    \dot{T}(t) = AT(t) - T(t)F(t) + BH(t),
\end{equation}
with any $T(0) \in \RR^{n_z \times n_x}$, in which
\begin{align}
\overline{z}(0)&= (T(0))^\oplus \overline x(0) - (T(0))^\ominus \underline x(0),\\
\underline{z}(0)& = (T(0))^\oplus \underline x(0) - (T(0))^\ominus \overline x(0).\label{eq:initzltv_c}
\end{align}
Similarly to the LTI case, we can show that~\eqref{eq:intobzltv_c}-\eqref{eq:initzltv_c} is an interval observer for system~\eqref{eq:syszltv_c}, so $\underline z (t) \leq z(t) \leq \overline z(t)$ for all $t \geq 0$. The Moore-Penrose inverse $(T(t))^\dagger$ of $T(t)$ is always defined but is not necessarily such that $(T(t))^\dagger T(t) = \Id$ for all $t \geq 0$. Consequently, while it is feasible to design an interval observer for system~\eqref{eq:syszltv_c} for all $t \geq 0$ in the $z$-coordinates, the technique used in Proposition~\ref{lem_linearcase} to recover the bounds of $x(t)$ for every $t \geq 0$ is not applicable here. Under Items~\ref{ass_fhltv_c} and~\ref{ass_obsltv_c} of Assumption~\ref{ass_sysltv_c} and with a right choice of the dynamics in the $z$-coordinates as detailed in Proposition~\ref{lem_ltv_c}, there exists a finite $t^\star \geq 0$ beyond which $T(t)$ becomes uniformly left-invertible and so $(T(t))^\dagger T(t) = \Id$ for all $t \geq t^\star$~\cite{bernardNonautonomousKKL}. This allows bringing back the bounds of~\eqref{eq:intobzltv_c}-\eqref{eq:initzltv_c} for system~\eqref{eq:sysxltv_c} after time $t^\star$, while no result is stated before this time. We then take for all $t\geq 0$,
\begin{align}
    \overline{x}(t) &= ((T(t))^\dagger)^\oplus \overline{z}(t) - ((T(t))^\dagger)^\ominus \underline{z}(t), \\
    \underline{x}(t) &= ((T(t))^\dagger)^\oplus \underline{z}(t)-((T(t))^\dagger)^\ominus \overline{z}(t),
    \end{align}
    \end{subequations}
which recover the bounds of $x(t)$ for all $t \geq t^\star$. Note that in this time-varying setting, the $z$-coordinates can have a higher dimension than the $x$-coordinates, so we typically cannot write the observer in the $x$-coordinates as in observer~\eqref{eq:obsxlin}. Such a design can still be done by adding fictitious states in the $x$-coordinates to equalize dimensions, but it has no clear interest compared to this. Our CT LTV design is recapped in the next proposition.
\begin{proposition}
    \label{lem_ltv_c}
    Suppose Assumption~\ref{ass_sysltv_c} holds and define $n_z = \sum_{i = 1}^{n_y} m_i$ (with $m_i$ coming from Item~\ref{ass_obsltv_c} of Assumption~\ref{ass_sysltv_c}). Consider any $T(0) \in \RR^{n_z \times n_x}$ and for each $i\in \{1, 2, \ldots,n_y\}$, a controllable pair $(\at_i, \bt_i)\in \RR^{m_i\times m_i}\times \RR^{m_i}$ with $\at_i$ Hurwitz and Metzler. 
There exists $\ell^\star > 0$ such that for any $\ell > \ell^\star$, there exist $t^\star \geq 0$ and $t \mapsto T(t)$ solution to~\eqref{eq:sylvesterltv_c} initialized as $T(0)$ with
\begin{subequations}
\begin{align}
 A &=\ell \at = \ell \diag(\at_1, \at_2, \ldots, \at_{n_y}) \in \RR^{n_z \times n_z}, \\ B& = \diag(\bt_1, \bt_2, \ldots, \bt_{n_y}) \in \RR^{n_z \times n_y},
\end{align}
\end{subequations}
that is uniformly left-invertible after time $t^\star$, i.e., there exists $c_T > 0$ such that $(T(t))^\top T(t) \geq c_T \Id$ for all $t \geq t^\star$. Then, observer~\eqref{eq:obszltv_c} initialized with that $T(0)$ is a finite-time interval observer for system~\eqref{eq:sysxltv_c}, i.e., we have $\underline{x}(t) \leq x(t) \leq \overline{x}(t)$ for all $t \geq t^\star$.
\end{proposition}
\textbf{Proof.} From~\cite[Theorem 4]{bernardNonautonomousKKL} which is the nonlinear version of our case, under Assumption~\ref{ass_sysltv_c} and with the chosen $(A,B)$, $t \mapsto T(t)$ taking the dynamics~\eqref{eq:sylltv_c} and initialized as $T(0)$ is solution to~\eqref{eq:sylvesterltv_c}, bounded, and uniformly left-invertible after a time, so $(T(t))^\dagger T(t) = \Id$ for all $t$ after this time. The rest is proven similarly to Proposition~\ref{lem_linearcase}, added with time variation of the matrices. \hfill $\blacksquare$

The following academic example illustrates our method.
\begin{example}
    Consider a controlled pendulum with a small angular position $\theta$, described by $\ddot{\theta}(t) + c\dot{\theta}(t) + \frac{g}{L(t)}\theta(t) = u(t)$
    where $c = 1$ is a damping coefficient, $g = 9.8$ is gravitational acceleration, and $L(t) = L_0 + L_1\sin(\omega t)$ for some $L_0 > 0$, $0 < L_1 < L_0$, and $\omega > 0$ is the pendulum's varying length, all in appropriate units. The known control is $u(t) = 0.15\sin(3t)$. We measure $\theta$. Picking $x_1 = \theta$ and $x_2 = \dot{\theta}$, we get the form~\eqref{eq:sysxltv_c} with $F(t) = \left(\begin{smallmatrix}
        0 & 1 \\ -\frac{g}{L(t)} & -c
    \end{smallmatrix}\right)$. With $g$ and $L(t)$ positive, $F(t)$ is not Metzler. Since this is time-varying, it is not obvious to find a transformation like~\cite{Mazenccontinu} into a Metzler form, but our method proves to be straightforward here. The pair $(F(t),H(t))$ is instantaneously observable with $m = 2$, and Item~\ref{ass_fhltv_c} of Assumption~\ref{ass_sysltv_c} is satisfied. Pick $\tilde{A} = \diag(-1,-2)$ and $\tilde{B} = (1,1)$, so $A = \ell \tilde{A}$ is both Hurwitz and Metzler, and then with $\ell = 2$, $T(t)$ is uniformly left-invertible after a time. Consider disturbance $d(t) = 0.04(\sin(1.2t),\sin(1.2t))$, noise $w(t) = 0.02\cos(20t)$ with unit gains, and assume known bounds for these. Observer~\eqref{eq:obszltv_c} is then designed for this system. Due to space constraints, we show in Figure~\ref{fig3} the results for only the second state, which is not measured. The interval observer is guaranteed after a certain time.
    \begin{figure}%[H]
        \centering
\includegraphics[width=0.34\textwidth,height=0.17\textwidth]{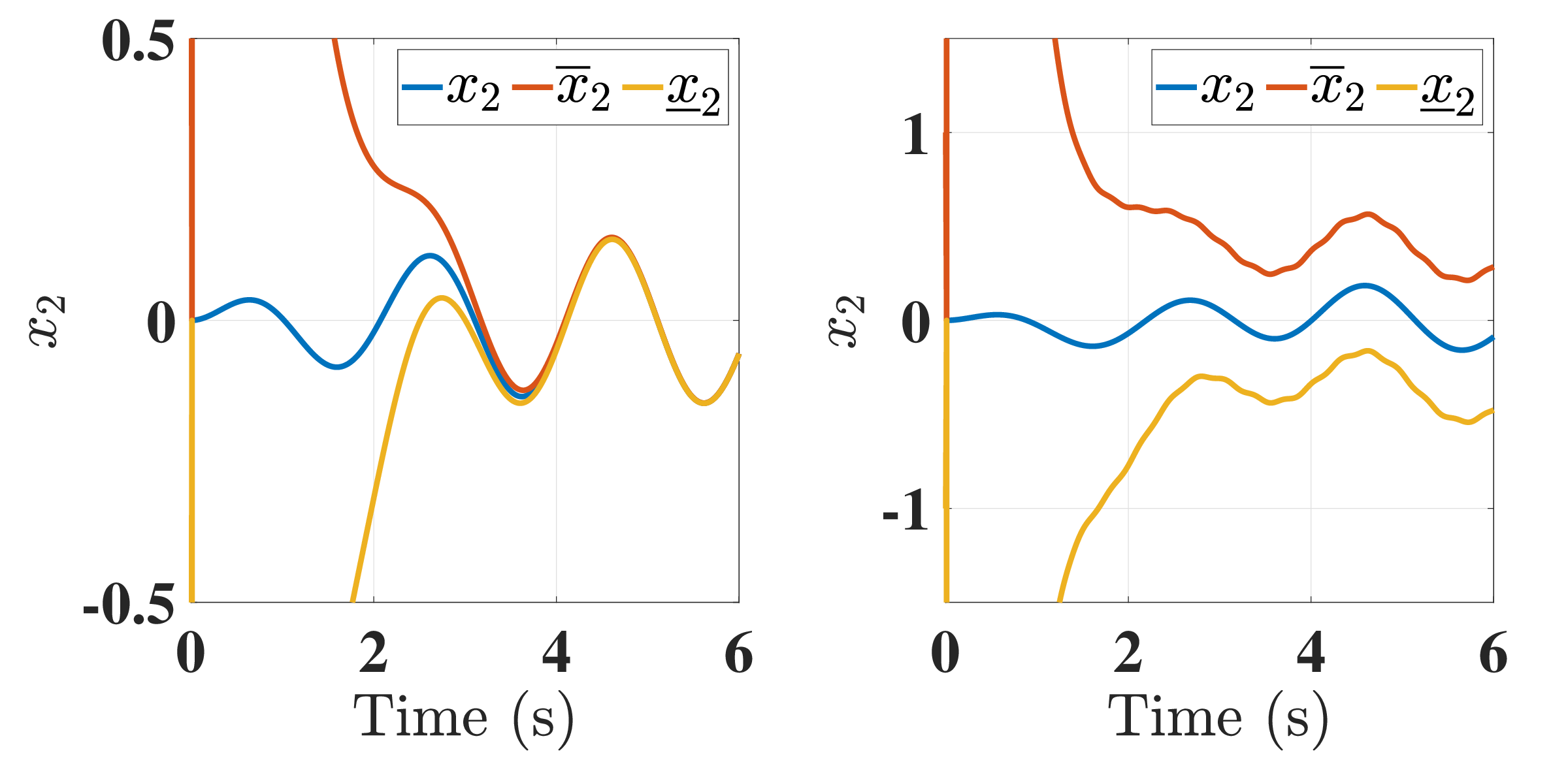}
        \caption{Finite-time estimation of $x_{2}$ in the absence and presence of disturbances$\slash$noise using observer~\eqref{eq:obszltv_c}.}       \label{fig3}
    \end{figure}
\end{example}

\subsection{Case of DT LTV systems}
In this section, consider a DT LTV system
\begin{equation}\label{eq:sysxltv_d}
        x_{k+1} = F_k x_k + u_k + D_kd_k, \quad
        y_k = H_k x_k + W_kw_k,
\end{equation}
with $(x_k,u_k,y_k,d_k,w_k)_{k \in \NN}$ defined in system~\eqref{eq:sysxltvintro} and $(F_k,H_k,D_k,W_k)_{k \in \NN}$ known matrices.
\begin{assumption}\label{ass_sysltv_d}
    For system~\eqref{eq:sysxltv_d}, we assume that:
\begin{enumerate}[label=(A3.\arabic*),leftmargin=*,nosep]
        \item \label{ass_invltv_d} For all $k \in \NN$, $F_k$ is invertible as $F_k^{-1}$;
        \item \label{ass_fhltv_d} There exist $c_f > 0$ and $c_h > 0$ such that for all $k \in \NN$, we have $\|F_k^{-1}\| \leq c_f$ and $\|H_k\| \leq c_h$;
        \item \label{ass_obsltv_d} The pair $(F_k,H_k)$ is uniformly completely observable (UCO),\footnote{This is the linear version of~\cite[Definition 1]{baokkltac} that is common in the Kalman literature, e.g.,~\cite{zhangKalman}, but here we take different $m_i$ for different output components instead of a single $m$ for all. In the KKL context, see for details in~\cite[Section VI.A]{baokkltac}.} i.e., there exist $c_o > 0$ and for each $i \in \{1,2,\ldots,n_y\}$, $m_i \in \NN$, such that for all $k \geq \max m_i$, we have
        $
\sum_{i = 1}^{n_y}\sum_{j = k-m_i}^{k-1}  (\OO_{i,j})^\top \OO_{i,j} \geq c_o \Id
$,
where $\OO_{i,j} = H_{i,j} F^{-1}_j \ldots F^{-1}_{k-2} F^{-1}_{k-1}$, with $H_{i,k}$ the $i^{\rm th}$ row of $H_k$;
        \item \label{ass_x0ltv_d} There exist $\underline x_0$ and $\overline x_0$ such that $\underline x_0 \leq x_0 \leq \overline x_0$;
        \item \label{ass_dwltv_d} There exist known sequences $(\overline d_k,\underline d_k,\overline w_k,\underline w_k)_{k \in \NN}$ such that $\underline d_k \leq d_k \leq \overline d_k$ and $\underline w_k \leq w_k \leq \overline w_k$ for all $k \in \NN$.
    \end{enumerate}
\end{assumption}
Following the KKL paradigm~\cite{baokkltac}, we strive for an LTV transformation $x_k \mapsto z_k := T_k x_k$, where $(T_k)_{k \in \NN}$ is a sequence of matrices satisfying
\begin{equation}\label{eq:sylvesterltv_d}
    T_{k+1}F_k = AT_k + BH_k, \quad \forall k \in \NN,
\end{equation}
where $A$ is Schur and non-negative, and $(A,B)$ is controllable, through which system~\eqref{eq:sysxltv_d} is put into
\begin{equation}\label{eq:syszltv_d}
        z_{k+1} = A z_k + B y_k + T_{k+1} u_k + T_{k+1} D_kd_k - BW_kw_k.
\end{equation}
We design an interval observer for system~\eqref{eq:syszltv_d}:
\begin{subequations}\label{eq:obszltv_d}
\begin{equation}
\left\{\begin{array}{@{}r@{\;}c@{\;}l@{}}\label{eq:intobzltv_d}
        \overline{z}_{k+1} &=&A \overline{z}_k +B y_k + T_{k+1} u_k  + (T_{k+1}D_k)^\oplus \overline d_k  \\&&{}-(T_{k+1}D_k)^\ominus \underline d_k + (BW_k)^\ominus \overline w_k - (BW_k)^\oplus \underline w_k\\
        \underline{z}_{k+1} &=& A \underline{z}_k + B y_k + T_{k+1} u_k +(T_{k+1}D_k)^\oplus \underline d_k \\&&{}-(T_{k+1}D_k)^\ominus \overline d_k + (BW_k)^\ominus \underline w_k - (BW_k)^\oplus \overline w_k,
    \end{array}\right.
\end{equation}
where the dynamics of $(T_k)_{k \in \NN}$ can be updated, under Item~\ref{ass_invltv_d} of Assumption~\ref{ass_sysltv_d}, as solution to~\eqref{eq:sylvesterltv_d} as
\begin{equation}\label{eq:sylltv_d}
    T_{k+1} = (AT_k + BH_k)F_k^{-1},
\end{equation}
with any $T_0 \in \RR^{n_z \times n_x}$, in which
\begin{equation}\label{eq:initzltv_d}
\overline{z}_0= T_0^\oplus \overline x_0 - T_0^\ominus \underline x_0,\quad
\underline{z}_0 = T_0^\oplus \underline x_0 - T_0^\ominus \overline x_0.
\end{equation}
Similarly to the LTI case, we can show that~\eqref{eq:intobzltv_d}-\eqref{eq:initzltv_d} is an interval observer for system~\eqref{eq:syszltv_d}, so $\underline z_k \leq z_k \leq \overline z_k$ for all $k \in \NN$. Under Items~\ref{ass_fhltv_d} and~\ref{ass_obsltv_d} of Assumption~\ref{ass_sysltv_d} and with a right choice of the dynamics in the $z$-coordinates as detailed in Proposition~\ref{lem_ltv_d}, there exists a finite $k^\star \in \NN$ linked to UCO beyond which the sequence $(T_k)_{k \in \NN_{\geq k^\star}}$ becomes uniformly left-invertible and so $T_k^\dagger T_k = \Id$ for all $k \in \NN_{\geq k^\star}$~\cite{baokkltac}. This allows bringing back the bounds of~\eqref{eq:intobzltv_d}-\eqref{eq:initzltv_d} for system~\eqref{eq:sysxltv_d} after time $k^\star$, stating nothing before. We take for all $k\in\NN$,
\begin{align}
    \overline{x}_k &= (T_k^\dagger)^\oplus \overline{z}_k - (T_k^\dagger)^\ominus \underline{z}_k, \\
    \underline{x}_k &= (T_k^\dagger)^\oplus \underline{z}_k-(T_k^\dagger)^\ominus \overline{z}_k,
    \end{align}
    \end{subequations}
which recover the bounds of $x_k$ for all $k \in \NN_{\geq k^\star}$. Our DT LTV design is recapped in the next proposition.
\begin{proposition}
    \label{lem_ltv_d}
    Suppose Assumption~\ref{ass_sysltv_d} holds and define $n_z = \sum_{i = 1}^{n_y} m_i$ (with $m_i$ coming from Item~\ref{ass_obsltv_d} of Assumption~\ref{ass_sysltv_d}). Consider any $T_0 \in \RR^{n_z \times n_x}$ and for each $i\in \{1, 2, \ldots,n_y\}$, a controllable pair $(\at_i, \bt_i)\in \RR^{m_i\times m_i}\times \RR^{m_i}$ with $\at_i$ Schur and non-negative. 
There exists $\gamma^\star \in (0,1]$ such that for any $0 < \gamma < \gamma^\star$, there exist $k^\star \in \NN$ and $(T_k)_{k\in\NN}$ where $T_k \in \RR^{n_z \times n_x}$ is solution to~\eqref{eq:sylvesterltv_d} initialized as $T_0$ with
\begin{subequations}
\begin{align}
 A &=\gamma \at = \gamma \diag(\at_1, \at_2, \ldots, \at_{n_y}) \in \RR^{n_z \times n_z}, \\ B& = \diag(\bt_1, \bt_2, \ldots, \bt_{n_y}) \in \RR^{n_z \times n_y},
\end{align}
\end{subequations}
that is uniformly left-invertible after time $k^\star$, i.e., there exists $c_T > 0$ such that $T_k^\top T_k \geq c_T \Id$ for all $k \in \NN_{\geq k^\star}$. Then, observer~\eqref{eq:obszltv_d} initialized with that $T_0$ is a finite-time interval observer for system~\eqref{eq:sysxltv_d}, i.e., we have $\underline{x}_k \leq x_k \leq \overline{x}_k$ for all $k \in \NN_{\geq k^\star}$.
\end{proposition}
\textbf{Proof.}
This is the discrete-time counterpart of the proof of Proposition~\ref{lem_ltv_c}, using~\cite[Theorems 2 and 3]{baokkltac}. \hfill $\blacksquare$

The following academic example illustrates our method.
\begin{example}
    Consider the system inspired from~\cite[Example 2]{zhangKalman}, with $F_k = \begin{pmatrix}
        1.2 & -1 + 0.5 \cos(k) \\ 0 & 0.5 + 0.2 \sin(k)
    \end{pmatrix}$ and $H_k = \begin{pmatrix}
        1 & 0
    \end{pmatrix}$. Here, $F_k$ is non-Schur and not non-negative, and it is not obvious to find classical transformations like~\cite{mazenc2014interval} for this time-varying system. Yet, our systematic KKL-based design proves to be handy in this case. From~\cite{zhangKalman}, the pair $(F_k,H_k)$ is UCO with $m = 2$, and Items~\ref{ass_invltv_d} and~\ref{ass_fhltv_d} of Assumption~\ref{ass_sysltv_d} are satisfied. Pick $\tilde{A} = \diag(0.1,0.2)$ and $\tilde{B} = (1,1)$, so $A = \gamma \tilde{A}$ is both Schur and non-negative, and then with $\gamma = 1$, $(T_k)_{k \in\NN}$ is uniformly left-invertible after $k^\star = 2$. Consider disturbance $d_k = 0.1(\sin(2k),-\sin(2k))$, noise $w_k = 0.02\cos(20k)$ with unit gains, and assume known bounds for these. Observer~\eqref{eq:obszltv_d} is then designed for this system. Due to space constraints, we show in Figure~\ref{fig2} the results for only the second state, which is not measured. The interval observer is guaranteed after $k^\star$.
    \begin{figure}%[H]
        \centering
\includegraphics[width=0.34\textwidth,height=0.17\textwidth]{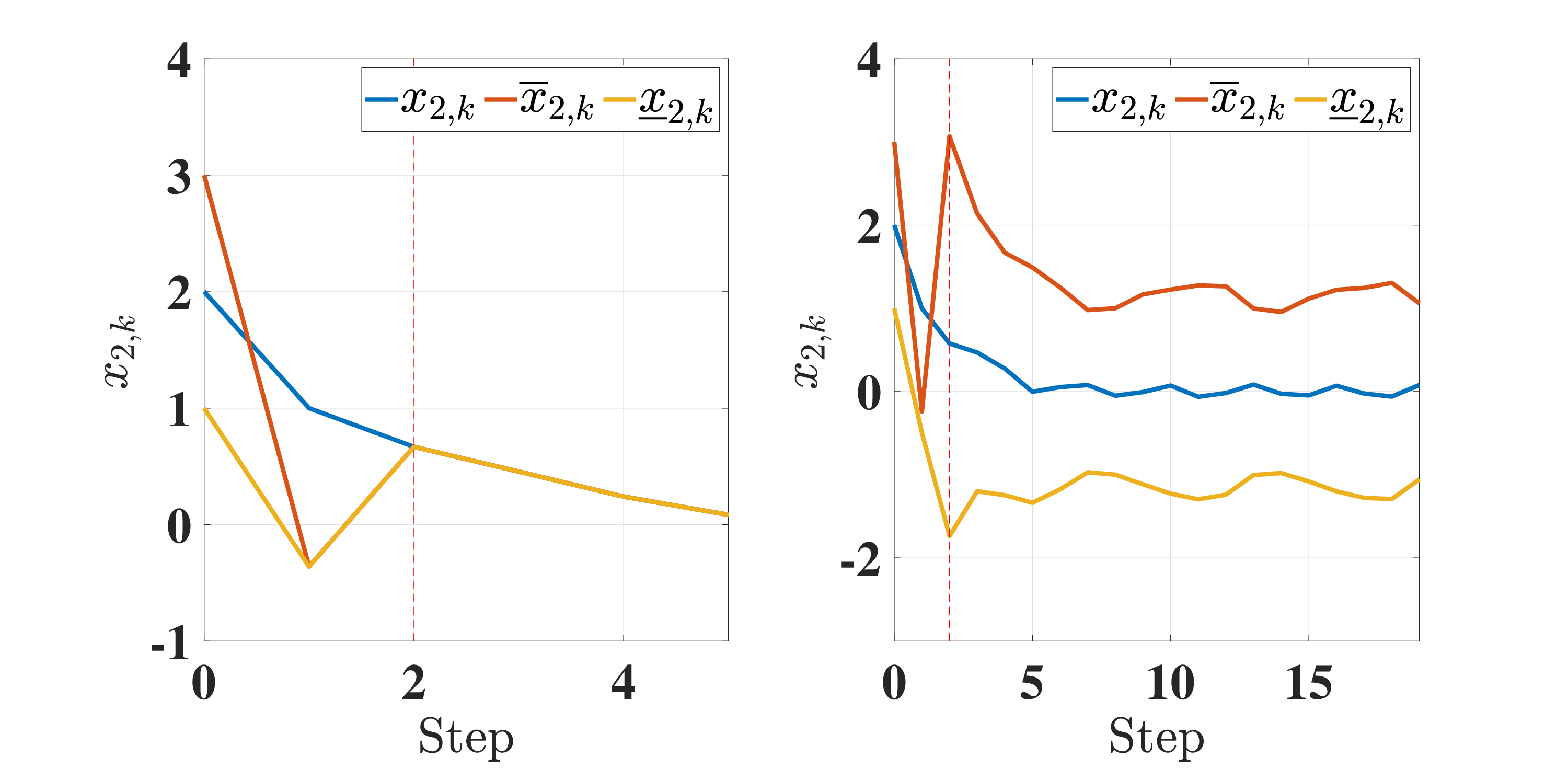}
        \caption{Finite-time estimation of $x_{2,k}$ in the absence and presence of disturbances$\slash$noise using observer~\eqref{eq:obszltv_d}.}       \label{fig2}
    \end{figure}
\end{example}

\section{Conclusion}
We introduce systematic interval observer designs for linear systems. Diverging from current methods, our approach boasts straightforward implementation and requires standard assumptions, with only a constant, easy-to-compute matrix $T$ to guarantee both positivity and stability for LTI interval observer design for an LTI system. %The interval observer design procedure for LTI systems can now be as straightforward as Algorithm~\ref{alg:cap}. 
Importantly, our method extends well to LTV systems. Notice that pushing $\ell$ in observer~\eqref{eq:obszltv_c} larger (resp., $\gamma$ in observer~\eqref{eq:obszltv_d} smaller) will speed up convergence rate but make us more sensitive to measurement noise~\cite{bernardNonautonomousKKL,baokkltac}. Future work is to effectively optimize the observer bounds through the choice of $(A,B)$.

\ack{We thank Pauline Bernard for her constructive remarks.}

\bibliographystyle{IEEEtran}       
\bibliography{ref}         

\begin{thebibliography}{10}
\providecommand{\url}[1]{#1}
\csname url@samestyle\endcsname
\providecommand{\newblock}{\relax}
\providecommand{\bibinfo}[2]{#2}
\providecommand{\BIBentrySTDinterwordspacing}{\spaceskip=0pt\relax}
\providecommand{\BIBentryALTinterwordstretchfactor}{4}
\providecommand{\BIBentryALTinterwordspacing}{\spaceskip=\fontdimen2\font plus
\BIBentryALTinterwordstretchfactor\fontdimen3\font minus
  \fontdimen4\font\relax}
\providecommand{\BIBforeignlanguage}[2]{{%
\expandafter\ifx\csname l@#1\endcsname\relax
\typeout{** WARNING: IEEEtran.bst: No hyphenation pattern has been}%
\typeout{** loaded for the language `#1'. Using the pattern for}%
\typeout{** the default language instead.}%
\else
\language=\csname l@#1\endcsname
\fi
#2}}
\providecommand{\BIBdecl}{\relax}
\BIBdecl

\bibitem{Mazenccontinu}
F.~Mazenc and O.~Bernard, ``{Interval observers for linear time-invariant
  systems with disturbances},'' \emph{Automatica}, vol.~47, no.~1, pp.
  140--147, 2011.

\bibitem{mazenc2014interval}
F.~Mazenc, T.~N. Dinh, and S.-I. Niculescu, ``{Interval observers for
  discrete-time systems},'' \emph{International Journal of Robust and Nonlinear
  Control}, vol.~24, no.~17, pp. 2867--2890, 2014.

\bibitem{Raissietal}
T.~Ra\"issi, D.~Efimov, and A.~Zolghadri, ``{Interval state estimation for a
  class of nonlinear systems},'' \emph{IEEE Transactions on Automatic Control},
  vol.~57, no.~1, pp. 260--265, 2012.

\bibitem{Efimov-12}
D.~Efimov, W.~Perruquetti, T.~Ra{\"\i}ssi, and A.~Zolghadri, ``{Interval
  observers for time-varying discrete-time systems},'' \emph{IEEE Transactions
  on Automatic Control}, vol.~58, no.~12, pp. 3218--3224, 2013.

\bibitem{luenberger}
D.~G. Luenberger, ``Observing the state of a linear system,'' \emph{IEEE
  Transactions on Military Electronics}, vol.~8, no.~2, pp. 74--80, 1964.

\bibitem{bernardNonautonomousKKL}
P.~Bernard and V.~Andrieu, ``{Luenberger observers for nonautonomous nonlinear
  systems},'' \emph{IEEE Transactions on Automatic Control}, vol.~64, no.~1,
  pp. 270--281, 2019.

\bibitem{baokkltac}
G.~Q.~B. Tran and P.~Bernard, ``{Arbitrarily fast robust KKL observer for
  nonlinear time-varying discrete systems},'' \emph{IEEE Transactions on
  Automatic Control}, vol.~69, no.~3, pp. 1520--1535, 2024.

\bibitem{gu}
D.-K. Gu, L.-W. Liu, and G.-R. Duan, ``Functional interval observer for the
  linear systems with disturbances,'' \emph{IET Control Theory \&
  Applications}, vol.~12, pp. 2562--2568, 2018.

\bibitem{franklin}
Z.~Zhang and J.~Shen, ``A survey on interval observer design using positive
  system approach,'' \emph{Franklin Open}, vol.~4, p. 100031, 2023.

\bibitem{hucheng}
Q.~Hu and D.~Cheng, ``{The polynomial solution to the Sylvester matrix
  equation},'' \emph{Applied Mathematics Letters}, vol.~19, no.~9, pp.
  859--864, 2006.

\bibitem{caronrichard}
R.~Caron and T.~Traynor, ``{The zero set of a polynomial},'' \emph{WMSR Report
  number 05-03}, 2005.

\bibitem{thachecc}
T.~N. Dinh, F.~Mazenc, and S.-I. Niculescu, ``Interval observer composed of
  observers for nonlinear systems,'' in \emph{{13th European Control
  Conference}}, 2014, pp. 660--665.

\bibitem{BriPetPra}
P.~Bristeau, N.~Petit, and L.~Praly, ``{Design of a navigation filter by
  analysis of local observability},'' in \emph{IEEE Conference on Decision and
  Control}, 2010, pp. 1298--1305.

\bibitem{zhangKalman}
Q.~Zhang, ``{On stability of the Kalman filter for discrete time output error
  systems},'' \emph{{Systems and Control Letters}}, vol. 107, pp. 84--91, 2017.

\end{thebibliography}
\end{document}